\documentclass[%
 preprint,
 superscriptaddress,
 amsmath,amssymb,
 aps,
 showkeys,
 titlepage,
 endfloats*.   
]{revtex4-2}

\usepackage[utf8]{inputenc}
\usepackage[english]{babel}

\usepackage{graphicx}
\usepackage{dcolumn}
\usepackage{bm}
\usepackage{booktabs}
\usepackage[colorlinks = true,
            linkcolor = blue,
            urlcolor  = blue,
            citecolor = red,
            anchorcolor = blue]{hyperref}

\begin{document}

\title[]{Unraveling the Surface Stability and Chemical Reactivity of Aza-Triphenylene Monolayer under O$_2$ and H$_2$O Exposure}

\author{Soumendra Kumar Das}
\affiliation {Department of Physics, Indian Institute of Technology (Indian School of Mines) Dhanbad, India}
\author{Prasanjit Samal}
\affiliation {School of Physical Sciences, National Institute of Science Education and Research (NISER) Bhubaneswar, HBNI, Jatni-752050, Odisha, India}
\author{Brahmananda Chakraborty}
\email{brahma@barc.gov.in}
\affiliation {High Pressure and Synchrotron Radiation Physics Division, Bhabha Atomic Research Centre (BARC), Trombay, Mumbai 400085, India}
\affiliation {Homi Bhabha National Institute, Anushaktinagar, Mumbai 400094, India}
\author{Sridhar Sahu}
\email{sridharsahu@iitism.ac.in}
\affiliation {Department of Physics, Indian Institute of Technology (Indian School of Mines) Dhanbad, India}

\date{\today}

\begin{abstract}
 Environmental oxidation has a great impact in tuning the physical, chemical and electronic properties of two-dimensional (2D) monolayers which can affect their practical applications in nanoscale engineering devices under ambient conditions. aza-triphenylene is a recently synthesized 2D materials whose practcal applications have not been systematically studied yet. In this study, we report for the first time, the adsorption and dissociation of O$_2$ and H$_2$O molecules on the surface of 2D aza-triphenylene monolayer through first principles calculations in combination with climbing image nudged elastic band (CINEB) method. The results indicates that both the O$_2$ and H$_2$O molecules weakly interact over the monolayer surface with an adsorption energy -0.16 eV and -0.37 eV respectively. In contrast, both the molecules exhibit resistance for dissociation due to the formation of energy barriers. The transition path indicates that molecular oxygen experience two energy barriers (0.16 ev and 1.22 eV) before getting dissociated atomic oxygen. However, the dissociation of  H$_2$O requires larger energy barrier (2.3 eV and 0.86 eV) due to breaking of covalent bonds and transfer of hydrogen. The strong chemical adsorption of atomic oxygen and H$^+$/OH$^-$ ions is due to the significant charge transfer from monolayer to the adsorbate as evidenced from the charge density difference and Bader charge analysis. Moreover, the dissociated configuration exhibit a larger band gap as compared to the pristine aza-triphenylene due to the strong hybridization between the p states of carbon and oxygen. our work predicts the robustness of azatriphylene monolayer against oxygen/water exposer thus ensuring their stability for device applications using these materials.

 
\end{abstract}

\maketitle

\section{\label{sec:level1}INTRODUCTION}
Two-dimensional (2D) materials have created considerable interest among the science community due to their fascinating electronic, physical, optical properties. Specifically, the discovery of graphene with its unique structural, electronic, and chemical properties and promising applications in electronics, optoelectronics, energy storage and conversion, light emitting devices, etc. along with its limitations of having zero band gap feature, provided a boost to search for other 2D materials beyond graphene with wide applications in various technological sectors\cite{bhimanapati2015recent}. This effort resulted in the exploration of a wide range of 2D materials such as carbides \cite{Anasori2017}, Carbonitrides \cite{zhang2024two}, hexagonal boron nitride \cite{chettri2021hexagonal}, graphyne \cite{kang2014oxygen}, holey-graphyne, BeN$_4$, transition metal dichalcogenides (TMDC) etc. Among these, aza-triphenylene monolayer is a recentely discovered 2D materials which is expected to be highly promising for energy applications such as oxygen evolution reaction (OER), electrochemical sensing etc. It is an interesting allotrope of carbon which has been experimentally synthesized via polycondensation reaction of 1,2,4,5-benzenetetramine tetrachloride and triquinoyl octahydrate. To the best of our knowledge, a very limited literature reports are available for its potential applications. Senthilkumar \textit{et al.} reported the hydrogen storage properties of Sc-decorated Aza COF through DFT calculations which predicted the stability of the sample upto 450K with hydrogen storage capacity of 8.43 wt \% of molecular hydrogen \cite{vasanthakannan2024hydrogen}. Amrutha   \textit{et al.} have studied the hydrgen storage application of Li decorated Aza covelent-organic-framework (COF) through first principles calculations and established that a hydrogen storage capacity of 9.49 wt\% can be achieved by decorating six Li atoms over the surface of Aza COF with each Li atom being capable to adsorb five H$_2$ molecules resulting in an average adsorption energy of -0.30 eV/H$_2$ \cite{amrutha2025designing}. First principles study performed by Kaur \textit{et al.} reported the suitablity of aza-triphenylene COF for efficient Na ion battries with a diffusion barrier of 0.78 eV for Na ion migration, specific heat capacity of 602.3 mAhg$^-1$, energy density of 1259.5 mWhg$^-1$ and a mean voltage of 0.62 V lying with the ideal range of 0 to 1 ev\cite{kaur2025aza}. In addition, DFt study predicted that the hydrogen storage performance of aza-triphenyle COF could be improved by surface functionalization with suitable elements like Zr, V, Ti\cite{tariq2025zr, mathew2025rational, amrutha2025enhancing}.

Various literature reports have established that the structural, chemical, and electronic properties of 2D materials undergo significant changes when exposed to atmosphere which can affect its device performance. The interaction with oxidizing species such as oxygen \cite{rawat2024first}, ozone \cite{das2024unveiling, patra2021ozonation}, water molecules \cite{patra2022surface}, and various other species can strongly affect the reactivity and stability of the sample. Rahman  \textit{et al.} have reported that the MoS$_2$ monolayer remains stable upto 1300 K when exposed to oxygen but at 1400 K temperature, oxygen starts reacting with it which became quite strong at a temperature of 1500 K \cite{rahman2021high}. Liu \textit{et al.} reported that the pristine TMDC ramains inert to oxygen adsorption whereas, oxidation occurs in presence of chalcogen vacancies with an adsorption energy varying between 1.8 eV to 2.9 eV \cite{liu2015atomistic}. The adsorption of oxygen and ozone molecules on the surface of 2D C$_2$N monolayer brought in substantial changes in its electronic structure with metallic state in physisorbed molecular configuration and an insulating state with a larger energy gap when dissociated to atomic oxygen. The oxygen molecule exhibited a very small energy barrier (0.05 eV) whereas the ozone molecule exhibited three energy barrier (0.56 eV, 0.53 eV and 0.43 eV) before getting dissociated to atomic oxygens \cite{das2024unveiling}. Abbasi \textit{et al.} reported that the adsorption of SO$_2$, SO$_3$ and O$_3$ molecules on the surface of MoS$_2$ monolayer exhibited energetically stable configuration with weak physisorption on the sample surface \cite{abbasi2019adsorption}. Combining experimental results and first principles study, Lee \textit{et al.} reported that the interaction of O$_3$ molecule on the surface of highly oriented pyrolytic graphite (HOPG) occurs through weak physisorption in molecular form whereas the dissociation to molecular oxygen and oxygen atom needs to overcome an energy barrier of 0.47 eV with a binding energy 0.72 eV \cite{lee2009ozone}. The oxygen adsorption  changes its electrnic properties of $\alpha$ - and $\beta$-graphyne from a metal to a semiconductor with spin splitting in $\gamma$-graphyne \cite{kang2014oxygen}. The first principles study on the adsorption and dissociation of H$_2$O molecules on the PuH$_2$ (110) surface indicates that H$_2$O in molecular form undergoes chemical adsorption on the PuH$_2$ surface with a charge transfer from the surface of Pu atoms to the O atoms in water molecules. The dissociation of H$_2$O involves  two energy barriers 0.406 eV and 0.718 eV before achieving the steady state accompanied by the twisting of the molecules \cite{luo2021adsorption}. By employing DFT analysis, Tran \textit{et al.} analysed the adsorption energy and dissociation barrier during the interaction of environmental oxygen, hydrogen and water molecules on the most relevant diamond surface C(001), C(110), C(111). The calculation revealed a strong correlation between the adsorption energy and surface energy and the dissociation process strongly alter the surface morphology and trobological properties of the diamond films \cite{tran2022ab}. Combining DFT and ab-initio atomistic thermodynamics method, Huang \textit{et al.} studied the adsorption of O$_2$, H$_2$ and H$_2$O molecules on the UO$_2$ (111) surface and reported a strong chemisorbed state for O$_2$ and H$_2$O adsorbed UO$_2$, whereas the H$_2$ undergoes weak physisorption without forming any chemical bonds\cite{huang2025chemisorption}. The interaction of O$_2$ and H$_2$O molecules on the ZrB$_2$ (0001) surface shown that O$_2$ exhibited stronger adsorption as compared to H$_2$O and the co-adsorption of both the molecules indicated that oxygen atoms adsorbed on the surface favors spontaneous dissociation of H$_2$O forming OH group and H$_2$ on the ZrB$_2$ surface \cite{Chen2025}. The Li ion storage capacity of graphatite based electodes can be enhanced by unctionalization of graphite using hydrogen-enriched water (HW) due to modulation in the charge distribution, lowering of diffusion barrier and enhancement of Li ion adsorption \cite{kim2025hydrogen}.

Realizing the importance of environmental oxidation on the electronic and physical properties of 2D materials, in this paper, we present for the first time, a comprehensive analysis on the adsorption and dissociation of O$_2$ and H$_2$O molecules on the surface stability and reactivity of aza-triphenylene monolayer using first principles density functional theory calculations at the PBE-GGA level. The electronic band structure calculations reveals a metallic state for the O$_2$ adsorbed and an insulating state for the dissociated configurations. The transition path during the dossociation of the molecules are analysed by climbing image nudged elastic band method (CINEB). The charge density and bader charge calculations illustrate a significant charge transfer for the dissociated configuration indicating a strong hybridization between the adsorbate and 2D monolayer.

\section{\label{sec:level2}Computational Methods}
First principles density functional theory (DFT) calculations were performed using the plane wave- pseudopotential code Vienna Ab initio Simulation Package (VASP)~\cite{kresse1996efficient,kresse1996efficiency}. The exchange and correlation functional were approximated by using the Perdew-Burke-Ernzerhof (PBE)~\cite{perdew1996generalized} parametrization-based generalized gradient approximation (GGA) method. The projected augmented wave type pseudopotentials were considered throughout the calculations. To accurately describe the long range interaction, we have adopted the DFT-D3 van der Waals corrections, developed by Grimme~\cite{grimme2011effect}. The crystal structure was generated using VESTA \cite{Momma2011}. A vacuum layer of 20 Å was added along the z direction to avoid interactions due to periodic boundary conditions. Initially, the O$_2$/H$_2$O molecule was kept at different possible sites at a distance of 4 Å from the surface and an ionic relax calculations were performed for all the configurations to get the optimized structure. The cut-off value for the kinetic energy was set at 520 eV. The convergence threshold for total energy and Helman-Feynman forces acting on each atom were set at 10$^{-6}$ eV and 0.01 eV/Å~ respectively. Brillouin zone integration were performed using the $\Gamma$-centred Monkhorst-Pack ($7 \times 7 \times 1$) \textbf{k}-mesh~\cite{monkhorst1976special} for the structural relaxation and ($9 \times 9 \times 1$) \textbf{k}-mesh for the electronic structure calculation of the pristine and adsorbed aza-triphenylene monolayer. However, a denser ($18 \times 18 \times 1$) \textbf{k}-mesh was considered for the calculations of the projected density of states using the tetrahedron method. The reaction mechanism for the adsorption and dissociation of O$_2$/H$_2$O molecule over the aza-triphenylene were analyzed by employing the climbing image nudged elastic band method ~\cite{henkelman2000improved}. The interaction of molecules and dissociated atoms/ions were studied by the charge density difference and Bader charge calculations ~\cite{bader1990atoms}.

The molecule aza-triphenylene interaction binding energies ($E_B$) have been computed as:
\begin{equation}
    E_B= E(molecule+aza)-E(molecule)-E(aza)
\end{equation}
where $E(molecule+aza)$, $E(molecule)$ and $E(aza)$ are the total energies of the molecule adsorbed aza-triphenylene configuration, the interacting molecule, and the pristine aza-triphenylene monolayer, respectively. The binding energies for the weakly adsorbed configurations are denoted as weak adsorption energy   ($E_{\text{physisorbed}}$). The difference in energies between the non-interacting configurations and the dissociated configurations is designated as dissociation energy ($E_{\text{diss}}$).

\section{\label{sec:level3}RESULTS AND DISCUSSION:}

\begin{figure}[tbp!]
\includegraphics[width=1.0\linewidth]{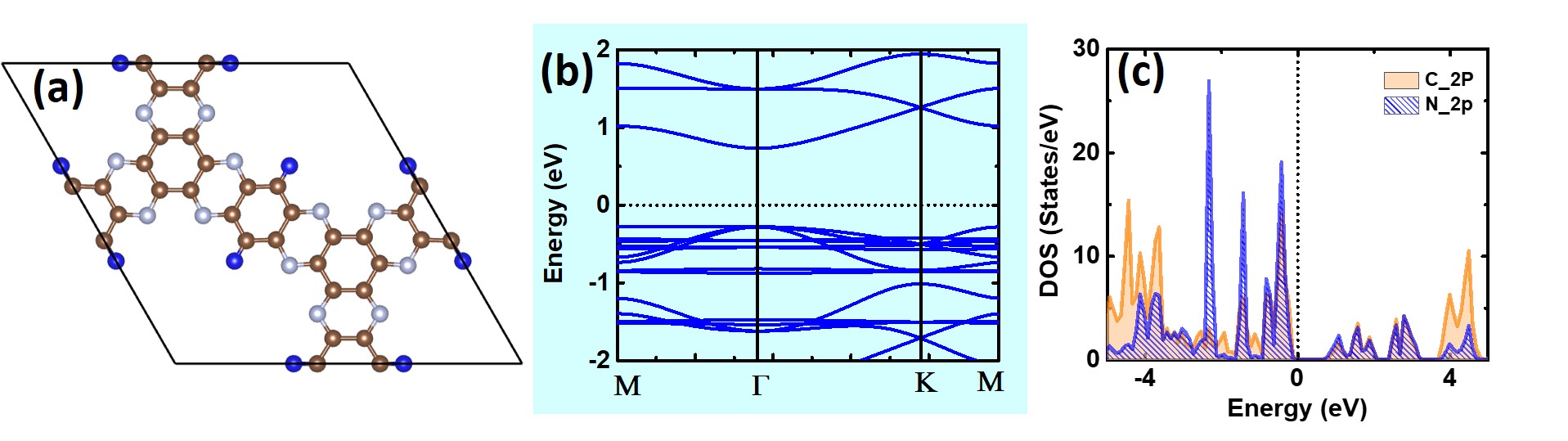}
\caption{\textbf{(a) Schematic of the crystal structure, (b) band structure and (c) partial density of states of pristine aza-triphenylene monolayer using PBE functional. In figure (a), the color code for the atoms are as follows: brown: C, grey: N, blue: H. Atomic sizes are exaggerated for clarity.
}}
         \label{fig:structure}
\end{figure} 

Figure~\ref{fig:structure} (a) indicates the schametic of the crystal structure of aza-triphenylene monolayer with single unit cell. The structure crystalizes in a hexagonal lattice with space group P1. It forms a covalent-organic framework (COF) consisting carbon, nitrogen and hydrogen atoms forming a porous structure. The pore size of the optimized unit cell is estimated to be 14.25 Å (see Figure S1 in the supplementary information (SI) file ) which is close to the reported experimental value of 14.1 Å \cite{liu2024recent}.   The single unit cell consists of 48 atoms in total out of which there are 30 carbon, 12 nitrogen and 6 hydrogen atoms arranged in such a way that each benzene ring is bridged by two or three pyrazine rings. Two hydrogen atoms are bonded to the central and edge sharing benzene rings respectively with C-H bond length 1.08 Å. After geometrical optimization, the lattice constant is estimated to be 16.54 Å which is consistent with the value 16.57 Å reported by Ghosh \textit{et al.} \cite{ghosh2024hot}. The C-C bond length for the central benzene ring varies between 1.45 Å (connected to pyrazine ring) and 1.40 Å (connected to hydrogen atoms). Similarly, the C-C bond length for the other benzene rings varies between 1.45 Å to 1.48 Å respectively. The twelve nitrogen atoms form six asymmetric pyrazine rings with C-N bond length varying between 1.32 Å to 1.36 Å. The calculated electronic band structure for the pristine aza-teiphenylene using PBE-GGA functional indicates a semiconducting nature with and energy gap of 1.1 eV which is consistent with other literature reports \cite{mathew2025rational, amrutha2025enhancing, kaur2025aza}. The valence band maximum (VBM) and conduction band minimum (CBM) occur at the $\Gamma$ point indicating its direct band gap nature (see Figure~\ref{fig:structure} (b)). The moderate band gap of pristine aza-triphenylene makes it promising for various applications such as hydrogen storage and catalysis. The calculated projected density of states for the pristine structure (see Figure~\ref{fig:structure} (c)) indicates that both the VBM and CBM are populated by equal contribution of '2p' states of carbon and nitrogen respectively.

\subsection{Structural Properties}

\begin{figure}[!tbp!]
    \includegraphics[width=1\linewidth]{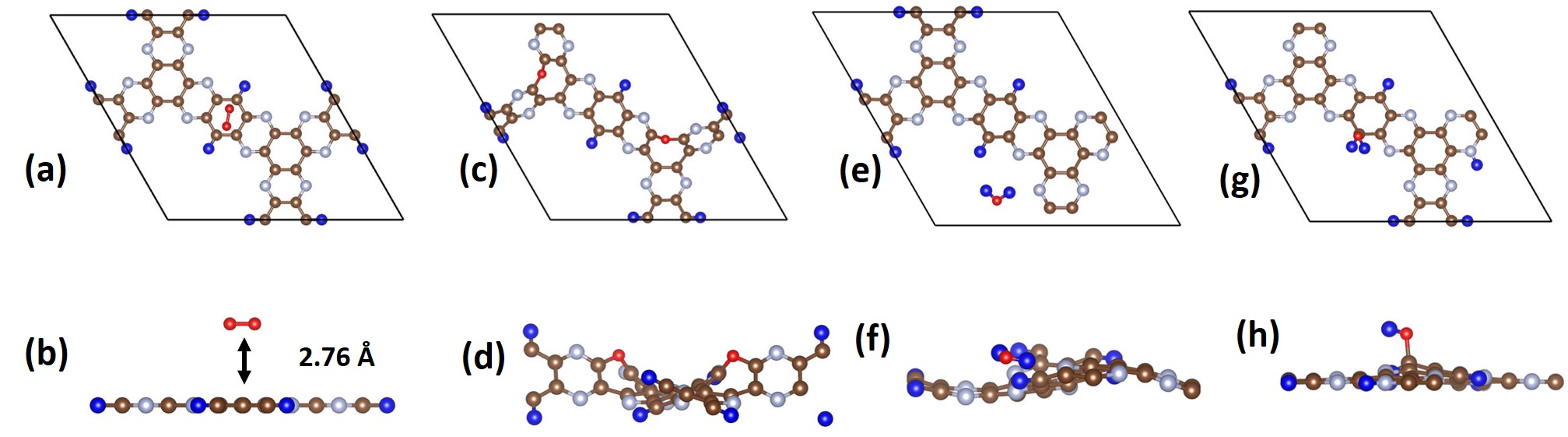}
         \caption{\textbf{Schematic representation of (a,b) O$_2$ adsorbed, (c,d) O$_2$ dissociated, (e,f) H$_2$O adsorbed , (g,h) H$_2$O dissociated aza-triphenylene monolayer. The color code for the atoms are as follows: brown: C, grey: N, blue: H, red: O. Atomic sizes are exaggerated for clarity.
}}
          \label{fig:oxygen_water_aza}
\end{figure} 

 In this section, we discuss the interaction of O$_2$ and H$_2$O molecules on the surface of aza-triphenylene monolayer before and after dissociation. In the begining, the O$_2$ and H$_2$O molecules were kept at different possible locations on the monolayer at a distance 4 Å from the surface and an ionic relax calculation was performed in each case to get the geometrically optimized structure (see Figure S2 in the supplementary information (SI) file ). The results established that the O$_2$ molecule prefers to stay horizontally over the central benzene ring containing two para hydrogen atoms and form the ground state configurations (Figure~\ref{fig:oxygen_water_aza}a). The surface to molecule distance decreases from 4 Å to 2.76 Å (Figure~\ref{fig:oxygen_water_aza}b) which is consistent with other literature reports for similar systems\cite{das2024unveiling}. This configuration describes a non-interacting state between the oxygen and sample known as weak-adsorption or Physisorption with a small adsorption energy of -0.16 eV (see Table S1 in the SI) . In contrast, when the molecule is dissociated into two atomic oxygens, they prefer to be chemisorbed on the aza-triphenylene surface on the two distant benzene rings forming C-O-C group (Figure~\ref{fig:oxygen_water_aza}c). The strong interaction of atomic oxygen with carbon distort the benzene ring by breaking the linear C-C bond and pushing the C-O-C bond towards the benzene ring with C-O bond length 1.38 Å (see (Figure~\ref{fig:oxygen_water_aza}d) and Table S1 in the SI). Similarly, the H$_2$O molecule after ionic relaxation at all possible locations prefers to stay in the hollow region of the aza-triphenylene surface (Figure~\ref{fig:oxygen_water_aza}e). Unlike the previous case with O$_2$ adsorption, the  H$_2$O stays almost close to the surface of the monolayer which produce a substantial distortion in its planar structure (Figure~\ref{fig:oxygen_water_aza}f). After dissociation of H$_2$O into H$^+$ and OH$^-$, the OH group forms chemical bond with the carbon atom of the central benzene ring whereas the H$^+$ ion moves to the edge of the unit cell and forms chemical bond with the N atom of the pyrazine ring ((Figure~\ref{fig:neb_o2_aza}g,h). The initial bond length between C-O was 3.27 Å which was significantly reduced to 1.45 Å after dissociation. The N-H bond length was calculated to be around 1.02 Å (see Table S1 in the SI).

\subsection{Oxygen reactivity}

\begin{figure*}[t!]
\includegraphics[width=1\linewidth]{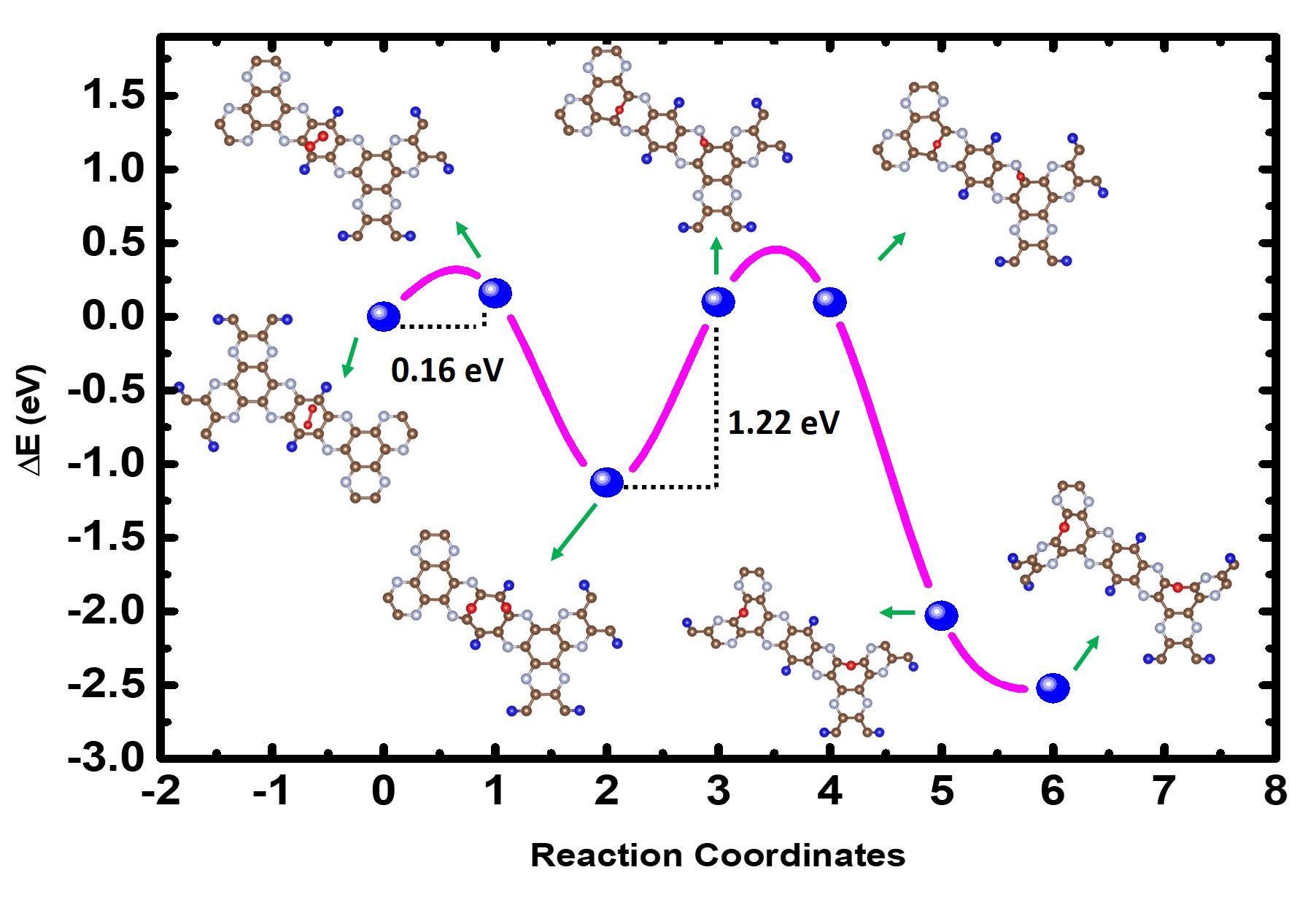}
\caption{\textbf{ Reaction pathway for O$_2$ with aza-triphenylene: Adsorption and dissociation process of O$_2$ on the aza-triphenylene monolayer surface, including structural configurations at intermediate stages. The colour scheme for atoms corresponds to that utilized in Fig.~\ref{fig:oxygen_water_aza}}}
         \label{fig:neb_o2_aza}
\end{figure*} 

In this section, we discuss the reaction mechanism of O$_2$ adsorption and dissociation on the surface of aza-triphenylene monolayer. Figure.~\ref{fig:neb_o2_aza} illustrates the transition path for the dissociation of O$_2$ molecule into atomic oxygen on the surface of aza-triphenylene monolayer which occurs via several intermediate steps. Initially the O$_2$ relaxes over the central benzene ring forming a non-interacting state with a small binding energy of -0.16 eV. The O-O bondlength after physirotpion is around 1.25 Å consistent with the bond length of free O$_2$ molecule. The Bader charge calculations and charge density difference plot also confirm this observation with a negligible charge (-0.07q) transfer from the C atom to the O atom (see Table S1 in SI). From the Figure~\ref{fig:neb_o2_aza}, it is clear that the dissociation of O$_2$ into atomic oxygen involves two energy barriers. The first energy barrier (0.16 eV) is due to the change in orientation of the O$_2$ from horizontal to a tilted configuration and movement from the center of the benzene ring to the edge. In the next step the two oxygen atoms were dissociated to atomic oxygen and forms chemical bonds with the carbon atoms of the same benzene ring. This results in a decrease in energy around 1 eV. In the next step, one oxygen atoms is transferred from the C-C bond of the benzene ring to the C-C bond of the adjacent pyrazine rings, while the other oxygen atom is bondd with the C-N atom of the pyrazine ring. This is a metastable state which involves breaking of C-O bond and subsequent crossing of the benzene ring, there fore exhibit a large energy barrier 1.22 eV. The fourth intermediate structure is quite similar to the third one with a very minimum change in energy. The transition to the fifth intermediate configuration involves  transfer of oxygen from the C-C bond and C-N bond of the pyrazine ring to the C-C bond of the adjacent benzene ring which is associated with a decrease in total energy by 1.93 eV. The final step involves a further decrease in total energy by 0.49 eV where the oxygen is bonded to the same C-C bond forming C-O-C group. However, the decrease in energy is due to the relaxation of the benzene and pyrazine rings which undergo change in orientation with respect to the plane of the surface (see Figure S3 in SI). In the final configuration, the oxygen atoms are more reactive and forms strong chemical bonds with the carbon with C-O bond length 1.38 Å approximately which is consistent with the C-O bond lenth with other oxidised 2D materials\cite{das2024unveiling}. 

\subsection{Water reactivity}

\begin{figure*}[t!]
\includegraphics[width=1\linewidth]{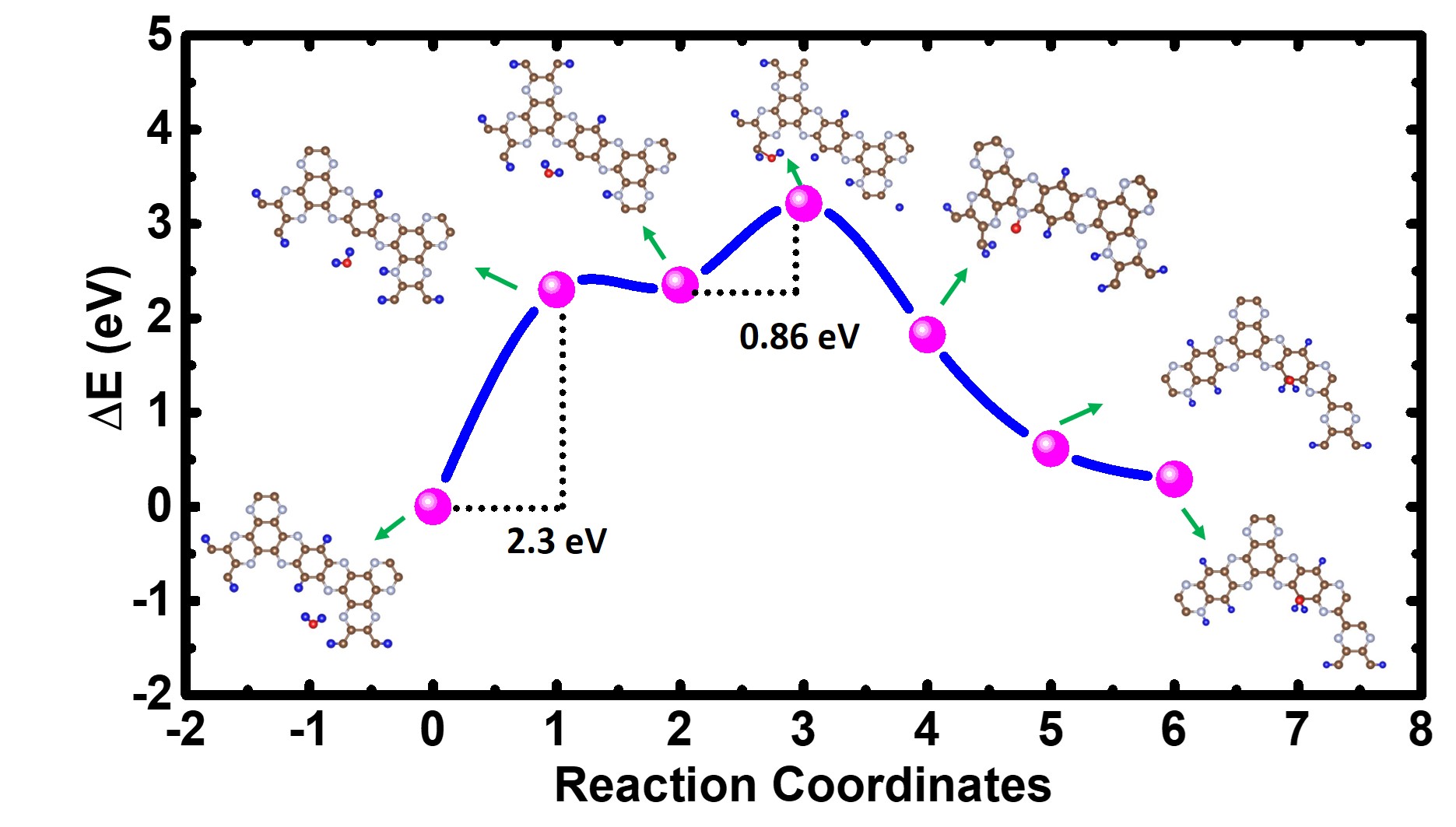}
\caption{\textbf{ Reaction pathway for H$_2$O with aza-triphenylene: Adsorption and dissociation process of H$_2$O on the aza-triphenylene monolayer surface, including structural configurations at intermediate stages. The colour scheme for atoms corresponds to that utilized in Fig.~\ref{fig:oxygen_water_aza}}}
         \label{fig:h2o_neb}
\end{figure*} 

In this section, we analyse the interaction of H$_2$O molecule and its dissociation on the surface of aza tri-phenylene monolayer. Figure.~\ref{fig:h2o_neb} shows the transition path of H$_2$O dissociation to H$^+$ and OH$^-$ calculated through CINEB approach. The initial configuration is determined by placing the H$_2$O molecule at a vertical distance of 4 Å from the surface of the monolayer at different possible locations and performing the ionic relax calculations to find the ground state configuration with minimum energy. It was found that the H$_2$O molecule prefer to adsorb weakly almost close to the hollow region of the surface. The binding energy for this adsorption forming the non-interacting state is around -0.37 eV (see Table S1 in the SI). The transition to the first intermediate metastable state requires to over come an energy barrier of 2.3 eV. Such a large value of energy barrier arrises due to breaking of C-H bond from the benzene ring and formation of N-H bond in the pyrazine ring. The second intermediate structure is similar to the previous one with minimal changes in the energy. The H$_2$O molecule which was parallel to the monolayer surface in the previous configuration now undergoes a little bit of tilting with respect to the xy plane. The transition to the next intermediate structure needs an energy barrier 0.86 eV due to breaking of H$_2$O into H$^+$ and OH$^-$. The H$^+$ ion is bonded to the C atom of the central benzene ring whereas the OH$^-$ ion moves to the cornor of the unit cell and forms chemical bonds with the C atom of the benzene ring near the edges. In the next intermediate configuration, the OH$^-$ ion bonded to the benzene ring at the edge, undergoes further dissociation where the H atom is chemically bonded to the C atom of the edge sharing benzene ring and the separated O atom forms bond with the N atom of the pyrazine ring. In the successive step, the N-H bond and N-O bond of the pyrazine breaks down and the detached H and O atom is connected to the C atom of the central benzene ring forming C-O-H group. This transition from the previous metastable state is associated with a decrease in total energy by 1.2 eV. In the final configuration, the chemical bonding of the OH$^-$ and H$^+$ remains unaltered. However a decrease in total energy (0.3 eV) is observed due to stabilization of the benzene and pyrazine rings from tilted configuration to the parallel orientation with respect to the ab plane. After dissociation of water, the C-O and N-H bond length in the corresponding benzene and pyrazine ring becomes 1.45 Å and 1.02 Å respectively.

\subsection{Charge density difference}

\begin{figure*}[t!]
\includegraphics[width=1\linewidth]{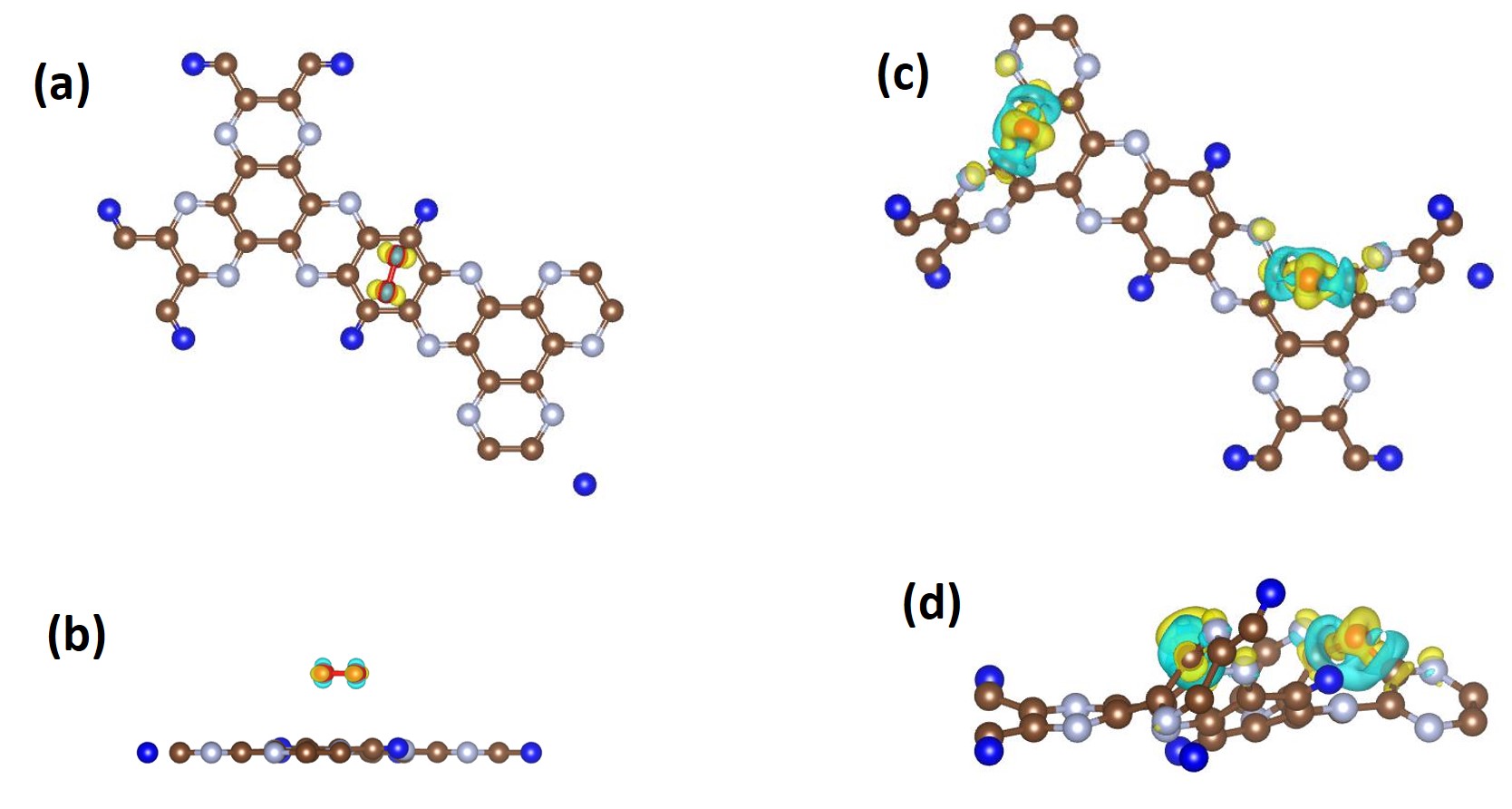}
\caption{\textbf{{Charge density difference of O$_2$ on aza-triphenylene: (a) physisorbed (O$_2$) and (b) dissociated (O + O) configurations. The iso-surface value was set at 0.05 e/Å$^3$. The yellow and cyan colours represent charge accumulation and depletion, respectively. The colour scheme for atoms corresponds to that utilized in Fig.~\ref{fig:oxygen_water_aza}.}}}
         \label{fig:chg_o2_aza}
\end{figure*}

\begin{figure*}[t!]
\includegraphics[width=1\linewidth]{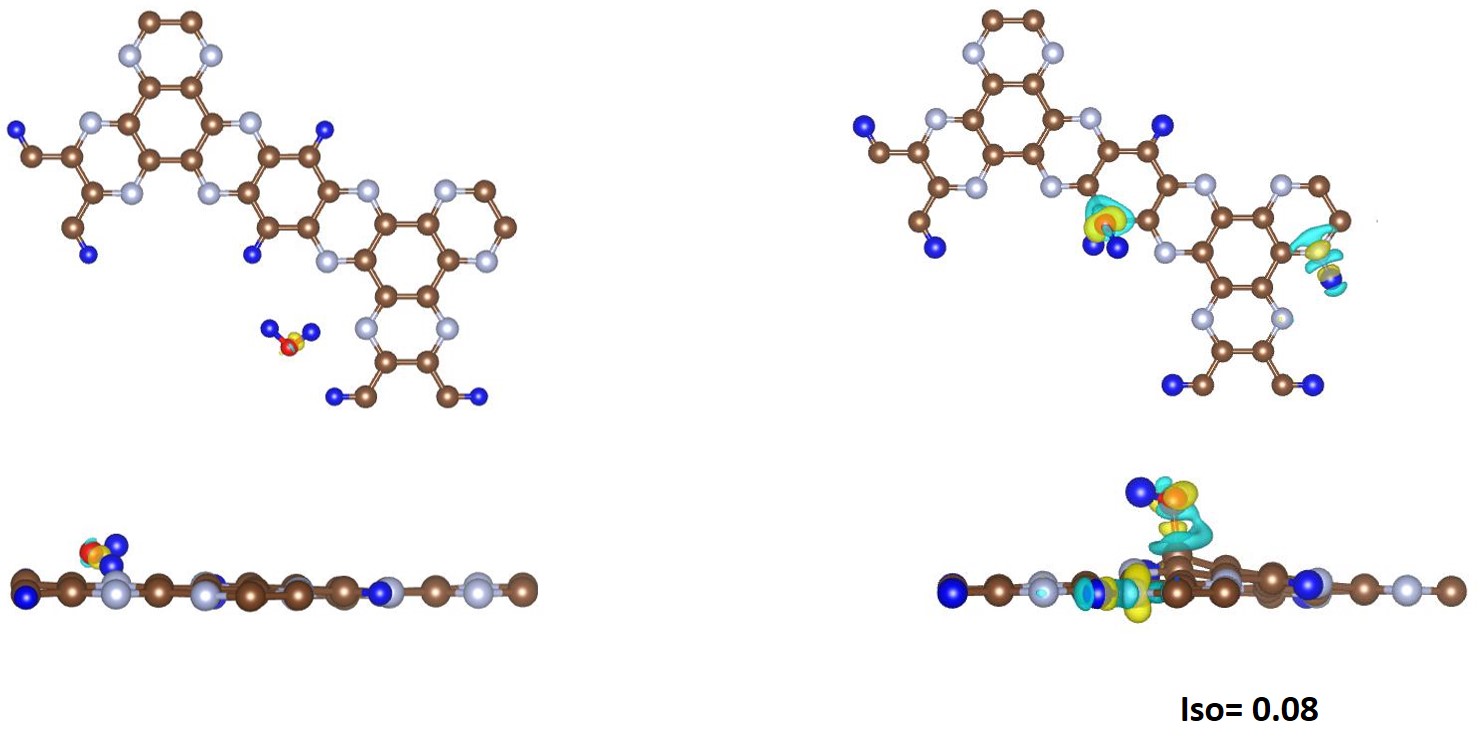}
\caption{\textbf{{Charge density difference of H$_2$O on aza-triphenylene: (a) physisorbed (H$_2$O ) and (b) dissociated (H$^+$ and OH$^-$) configurations. The iso-surface value was set at 0.05 e/Å$^3$. The yellow and cyan colours represent charge accumulation and depletion, respectively. The colour scheme for atoms corresponds to that utilized in Fig.~\ref{fig:oxygen_water_aza}.}}}
         \label{fig:chg_h2o_aza}
\end{figure*} 

The reaction mechanism for the adsorption and dossiciation of O$_2$ and H$_@$O is further supportted by the charge density difference and bader charge calculations. Figure.~\ref{fig:chg_o2_aza} (a,b) indicates the charge density difference plot for the weakly adsorbed O$_2$ on the surface of aza-triphenylene. We would like to mention here that in all the charge density difference plots, the cyan indicates a charge depletion and yellow color indicates a charge accumulation. The figure clearly indicates that a very negligible charge appears on the oxygen molecule. In contrast for the dissociated configuration (Figure.~\ref{fig:chg_o2_aza} c,d), the cyan color appears near the carbon (charge loss) and yellow color appears near the oxygen atoms (charge gain). This is further supported by the Bader charge calculations. From table S1, we found that a very small negative charge (-0.07q) appears near the oxygen molecule with a corresponding small positive charge (+0.04q) near the carbon atom. On the otherhand, the for the dissociated configuration, a large negative charge (-1.77q) is accumulated over the oxygen and a corresponding positive charge (+2.51q) on the carbon atom. All these observations confirm a charge transfer from carbon to oxygen. 

Similarly, for the H$_2$O adsorption and dissociation, the charge density difference plot (see (Figure.~\ref{fig:chg_h2o_aza} a,b) indicates a  charge accumulation for the physisorbed configuration. The corresponding Bader charge calcultions estimates a negative charge (-2.03 q) on the oxygen atom and +1.45q on the carbon atom. After dissociation (see (Figure.~\ref{fig:chg_h2o_aza} c,d) a strong depletion of charge is observed on the carbon atom of the central benzene ring and N atom of the pyraine ring. Similarly, a yellow color appears on the oxygen atom indicating a chatge accumulaion. Bader charge calculations predict a negative charge (-1.69q) on oxygen and positive charge (+0.75 q) on the carbon and nitrogen atoms. Thus there is a strong charge transfer from C/N to oxygen and hydrogen during the dissociation of H$_2$O.

\subsection{Electronic structure analysis}

\begin{figure*}[t!]
\includegraphics[width=1\linewidth]{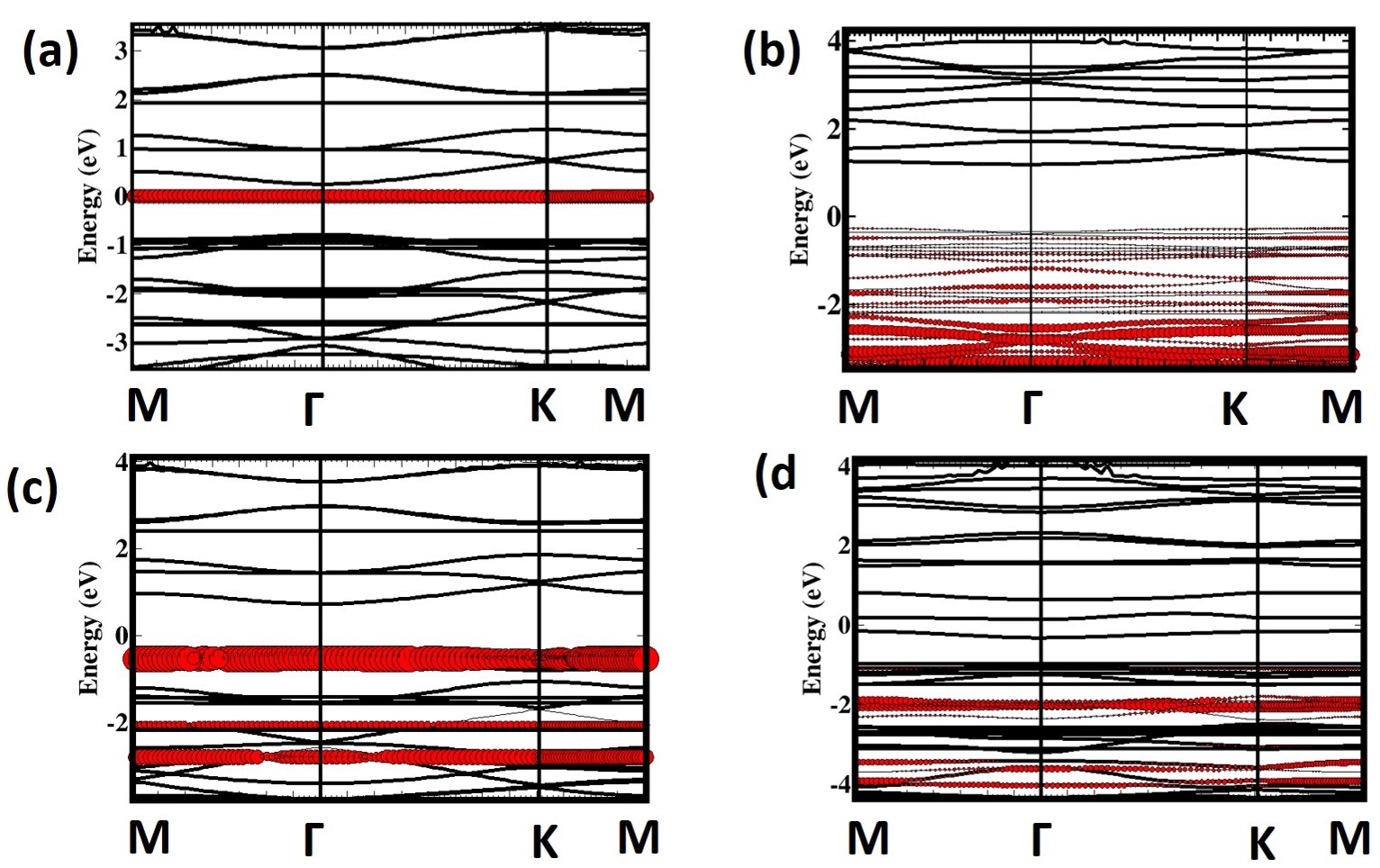}
\caption{\textbf{Orbital Projected Electronic Structure for O$_2$/H$_2$O interaction with aza-triphenylene}: Top panel: (a) physisorbed (O$_2$), and (b) dissociated (O + O) oxygen configurations. Bottom panel: (c) physisorbed (H$_2$O) and (d) dissociated ( H$^+$ and OH$^-$) water configurations.}
         \label{fig:proj_band}
\end{figure*} 

In this section we discuss the electronic properties of 0$_2$/H$_2$O adsorbed aza tri phenylene monolayer in their physisorbed and dissociated configuration through the orbital projected band structure analysis. The 0$_2$ molecule forms a non-interacting state by weakly adsorbed on the surface the system and exhibit a metallic behaviour which originated from the oxygen '2p' states dispersing at the Fermi level (Figure~\ref{fig:proj_band}a). However, after dissociating into atomic oxygen (Figure~\ref{fig:proj_band}b), the monolayer exhibit a wider band gap as compared to the pristine  aza-triphenylene. The oxygen '2p' states are found to be dispersed deep inside the valence band around 2eV below the Fermi level. This indicates a strong hybridization between the '2p' states of carbon and oxygen. Unlike the case of 0$_2$, when H$_2$O molecule is adsorbed in the hollow region of the aza tri phenylene it forms an insulating state (Figure~\ref{fig:proj_band}c). The oxygen '2p' states are dispersed below the Fermi level without crossing it. The direct band gap nature remains unaffected during the adsorption. However, when H$_2$O is dissociated into H$^+$ and OH$^-$ ion, the system exhibit a nearly metallic state with the valence band maximum (VBM) and conduction band minimum (CBM) overlapping with each other very close to the Fermi level (Figure~\ref{fig:proj_band}d). It is interesting to note that this narrow band gap state is not due to the oxygen interaction as the '2p' states of oxygen lie around 2eV below the Fermi level.

\begin{figure*}[t!]
\includegraphics[width=1\linewidth]{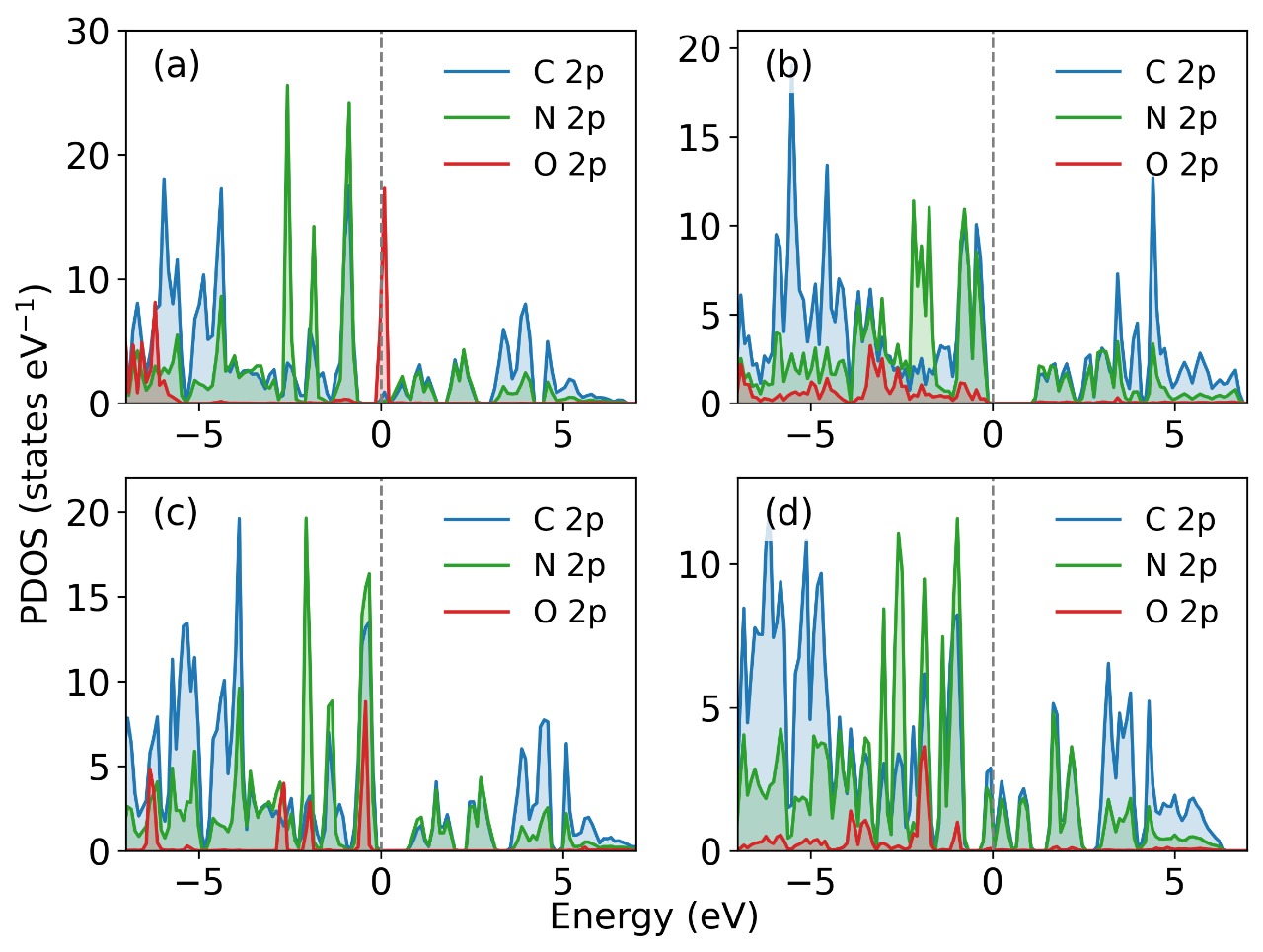}
\caption{\textbf{Orbital Projected Density of States for O$_2$/H$_2$O interaction with aza-triphenylene}: Top panel: (a) physisorbed (O$_2$), and (b) dissociated (O + O) oxygen configurations. Bottom panel: (c) physisorbed (H$_2$O) and (d) dissociated ( H$^+$ and OH$^-$) water configurations.}
         \label{fig:dos}
\end{figure*}

To provide a deeper analysis for the formation of metallic and semiconducting state, we performed projected density of states calculations for the adsorbed and dissociated structures which is illustrated in Figure~\ref{fig:dos}. For the physisorbed of O$_2$ molecule on the monolayer, the oxygen 2p orbital forms an impurity state which is highly localized at the Fermi level (Figure~\ref{fig:dos}a). The states appearing below the Fermi level is largely populated by the '2p' states of nitrogen whereas the empty states appearing in the conduction band region near the Fermi level have equal population of nitrogen and carbon '2p' states. In contrast, after dissociation to atomic oxygen, the 2p states of oxygen are populated far from the Fermi level in the valence band region (Figure~\ref{fig:dos}b). The carbon and nitrogen '2p' states forms the valence band maximum and conduction band minimum with nearly equal population. However their orbital occupancy becomes larger as compared to the physisorbed configuration. Similarly, during the H$_2$O adsorption in the hollow region, the VBM is populated largely by nitrogen and oxygen '2p' states and the CBM lies 0.5 eV above the Fermi level (Figure~\ref{fig:dos}c). Interestingly, after dissociation, the CBM crosses the Fermi level and forms the metallic state where as the oxygen '2p' states are dispersed in the valence band region (Figure~\ref{fig:dos}d).

\subsection{Reduced Density Gradient Analysis}

\begin{figure*}[t!]
\includegraphics[width=1\linewidth]{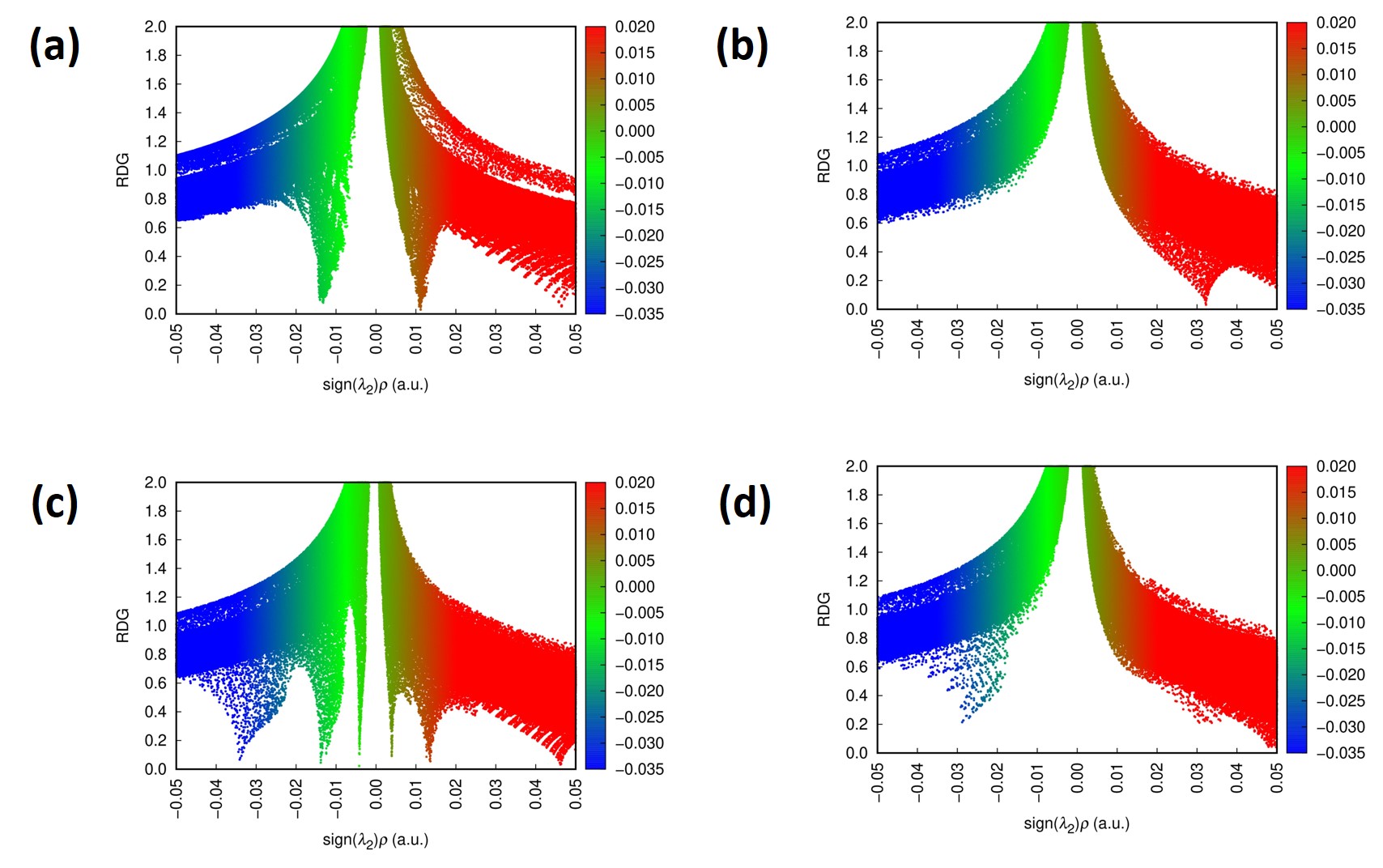}
\caption{\textbf{Reduced density gradient (RDG)
plot for the adsorption and dissociation of O$_2$ and H$_2$O in aza-triphenylene monolayer}: Top panel: (a) physisorbed (O$_2$), and (b) dissociated (O + O) oxygen configurations. Bottom panel: (c) physisorbed (H$_2$O) and (d) dissociated ( H$^+$ and OH$^-$) water configurations.}
         \label{fig:RDG}
\end{figure*}

In order to get a deep insight into the adsorption and dissociation of O$_2$ and H$_2$O molecules on the surface of aza-triphenylene monolayer, we have performed the reduced density gradient (RDG) calculation (see Figure~\ref{fig:RDG}) which provides a visual representation of the nature and strength of the molecular interactions. In the horizontal axis, the positive and negative values of $\mathrm{sign}(\lambda_2)\rho$ indicates the repulsive and attractive interactions respectively. The vertical axis represents the magnitude of the interaction. The appearance of red and green color near the zero of the x-axis indicates the steric repulsive interaction and weak van der Waals interaction where as the blue color indicates the strong attractive interactions. For the O$_2$ adsorbed aza-triphenylene configuration (see Figure~\ref{fig:RDG}a), the RDG plot illustates dominants green regions near the zero of $\mathrm{sign}(\lambda_2)\rho$ suggesting weak non-covalent van der Waals interactions. After the dissociation into atomic oxygen, robust blue regions appear at negative $\mathrm{sign}(\lambda_2)\rho$, indicating the strong attractive interactions and bond formation between atomic oxygen and monolayer surface (Figure~\ref{fig:RDG}b). Similarly, during the weak physical adsorption of H$_2$O molecule on the monolayer surface, the RDG plot exhibits strong green regions near the zero of $\mathrm{sign}(\lambda_2)\rho$ indicating the weak hydrogen bonding or physisorption between H$_2$O and monolayer surface (Figure~\ref{fig:RDG}c). For the H$_2$O dissocation to H$^+$ and OH$^-$, the RDG plot  (see Figure~\ref{fig:RDG}d) emerges with distince blue regions at negative $\mathrm{sign}(\lambda_2)\rho$ indicating strong local attractive interaction betweeen the adsorbates and monolayer confirming the presence of chemisorption.

\section{\label{sec:level4} CONCLUSION:}
In colclusion, we report the reactivity of O$_2$ and H$_2$O molecules on the surface of aza-triphenylene monolayer under ambient conditions through first principles DFT calcilations. The O$_2$ molecule gets physically adsorbed at 2.74 Å above the central benzene ring forming a non interacting state. After dissociation into atomic oxygen it forms strong chemical bond forming C-O-C group with the benzene ring. The transition path during the adsorption and dissociation of O$_2$ into atomic oxygen involves two energy barriers 0.12 eV and 1.22 eV respectively due to breaking of chemical bonds as evident from the CINEB calculations. Simillarly, the H$_2$O molecule is adsorbed in the hollow region of the aza-trphenylene lying very close to the surface. After dissociation to OH$^-$ and H$^+$ ion, it exhibits chemisorption state by forming bonds with cabron and nitrogen atoms respectively. The corresponding NEB result indicates two large energy barriers of 2.3 eV (due to breaking of C-H bond of the benzene ring) and 0.86 eV due to the dissociation of H$_2$O. The charge density difference in combination with Bader charge indicates a strong charge transfer for the dissociated configuration. The electronic structure calculations indicate a larger energy gap for the oxygen dissociated structure as compared to the pristine monolayer. The corresponding partial density of states reveals that the oxygen '2p' states lie exactly at the Fermi level for the physisorbed state and dispersed deep inside the valence band for the dissociated configuration thus indicating a strong hybridization between carbon and oxygen atoms. The present study provides evidence for the stability of the aza-triphenylene monolayer upon environmental oxidation and hydrogenetion and shed light on the suitability for device applications in various technological domains.

\begin{acknowledgments}
The authors acknowledge the use of the high-performance computing facility at  National Institute of Science Education and Reserch (NISER), India.
\end{acknowledgments}

\section*{Conflict of interest}
The authors have no conflicts to disclose.

\section*{Supporting Information}

The additional data that support the findings of this article are available in the Supplementary Information (SI).

\nocite{*}
\bibliography{ref}

\end{document}